\documentclass[12pt]{article} 

\usepackage{mathtools, amsfonts, amsthm, latexsym, amssymb,dsfont}
\usepackage[T1]{fontenc}
\usepackage[tracking=true]{microtype}
\usepackage{lmodern}
\usepackage{verbatim}

\usepackage[margin=1in]{geometry}

\usepackage{verbatim}
\usepackage{graphicx}
\usepackage{subcaption}
\usepackage{booktabs}
\usepackage{longtable}
\usepackage{cite}

\usepackage{color}
\definecolor{darkblue}{cmyk}{0.9,0.9,0,0}
\definecolor{wine-stain}{rgb}{0.5,0,0}
\usepackage[colorlinks, allcolors=darkblue, linktoc=all]{hyperref} 
\usepackage{setspace}

\renewcommand{\L}{\mathcal L}

\newcommand{\beq}{\begin{equation}}
\newcommand{\eeq}{\end{equation}}
\newcommand{\be}{\begin{equation}}
\newcommand{\ee}{\end{equation}}
\newcommand{\bea}{\begin{align}}
\newcommand{\eea}{\end{align}}

\def\gsim{ \lower .75ex \hbox{$\sim$} \llap{\raise .27ex \hbox{$>$}} }
\def\lsim{ \lower .75ex \hbox{$\sim$} \llap{\raise .27ex \hbox{$<$}} }

\def\e{{\epsilon}}

\def\g {{\gamma}}
\def\beq{\begin{eqnarray}}
\def\eeq{\end{eqnarray}}

\def\mpl{M_{\rm Pl}}
\def\e{{\epsilon}}

\def\s{\sigma}

\def\g{{\cal G}}

\def\L*{{\cal L}_*}
\def\L{\mathcal{L}}
\def\({\left(}
\def\){\right)}

\def\<{\langle}
\def\>{\rangle}

\def\g{\bar g}
\def\R{\bar R}
\def\hg{\hat g}
\def\M{\bar M}
\def\n{\nabla}
\def\hn{\hat \nabla}

\def\xyma{\xymatrix@M.7em}
\def\xymas{\xymatrix@M.1em}
\newcommand{\ba}{\begin{eqnarray}}
\newcommand{\ea}{\end{eqnarray}}

\begin{document}
\thispagestyle{empty}
\vspace*{0.5in} 
\begin{center}
{\Large \bf A New Gravitational Action For The Trace Anomaly}
\vskip 0.25cm

\vskip 0.7cm
\centerline{
{\large {Gregory Gabadadze}}
}
\vspace{.2in} 

\centerline{{\it Center for Cosmology and Particle Physics, Department of Physics}}
\centerline{{\it New York University, 726 Broadway, New York, NY, 10003, USA}}
\end{center}

\begin{abstract}
The question of building a local diff-invariant effective gravitational action for the 
trace anomaly is reconsidered. General Relativity (GR) combined with the existing action for 
the trace anomaly is an inconsistent low energy effective field theory. 
This issue is addressed by extending GR into a certain scalar-tensor theory, which 
preserves the GR trace anomaly equation, up to higher order corrections. 
The extension introduces a new mass scale  -- assumed to be below the Planck scale --
that governs four high dimensional terms in a local diff-invariant trace 
anomaly action. Such terms can be kept, while an infinite number 
of Planck-suppressed invariants are neglected. 
The resulting theory maintains two derivative equations of motion. In a certain 
approximation it reduces to the conformal Gallileon, which could have physical  
consequences.

\end{abstract}

\newpage
\noindent

\section{Introduction and summary}

General Relativity (GR), combined with a  
quantum field theory exhibiting the gravitational trace anomaly \cite {Duff}, and 
described by its effective action  \cite {Riegert,FT}, is an inconsistent field theory, despite
the existence of a local diff-invariant trace anomaly action 
(see, \cite {KS,GGTukhashvili}, and discussions below).  
Such a theory is strongly coupled at an arbitrarily 
low energy scale, as was shown in \cite {Kurt} in a different context.  
The inconsistency stems from the fact that  the conformal part of a metric is not a 
propagating degree of freedom in classical GR\footnote{A propagating degree of freedom  
is defined as a mode with proper quadratic time derivative and spatial gradient terms in Minkowski 
space-time, which is not removable by gauge transformations or by field redefinitions, 
nor it is restricted by constraints. The scale factor  of the Fridmann-Lemaitre-Robertson-Walker (FLRW)   metric  
is a conformal mode which has a ghost-like ("wrong sign") quadratic time derivative term in 
the GR action. This "wrong-sign" is crucial for FLRW cosmology. However, the scale factor is 
not a propagating degree of freedom because its spatial dynamics is constrained within GR; 
had it been  a propagating mode, it would have led to  unsurmountable 
instabilities because of its ghost-like quadratic time derivative term. The GR action supports only two 
propagating modes, the helicity $\pm 2$ states of a massless graviton, often referred as the tensor modes.}. 
This issue is fully relevant to GR coupled to the Standard Model (SM) of particle physics. 

One way to avoid the inconsistency is to augment GR so that the  conformal mode turns 
into a proper propagating degree of freedom, without spoiling observational consequences of GR, 
as it is proposed in this work. An alternative to the augmentation of GR would be to cancel the trace anomaly 
by introducing  additional low energy degrees of freedom. 

As a reminder, one is concerned with the invariance of a 
quantum field theory (QFT) -- such as the SM  -- with respect to global scale transformations. 
At the classical level the scale invariance could be exact, 
as in the Maxwell theory, or it could be violated explicitly, as in a massive scalar theory,
or in GR. The quantum theory violates the scale invariance generically, irrespective of whether the classical theory is 
or is not scale invariant.  This violation  appears in the trace of a stress-tensor, as it was first
shown for gauge  fields \cite {Crewther,ChanowitzEllis}, and subsequently  for a 
gravitational field \cite {Duff}. The latter will be the  focus on this work. 

It is useful to build a low energy effective  action that would incorporate quantum loops of the 
matter fields (see, \cite {BirDav} and references therein). The variation of such a quantum effective 
action would give rise to the equations of motion which capture the trace anomaly \cite {Riegert,FT}. 
These equations could then be solved in various physical settings, notably in astrophysics and cosmology,
with a benefit that the solutions would automatically contain  
quantum effects due to the trace anomaly. 

Riegert  constructed a local effective action, $S_A$,  which captures the trace anomaly 
in its equations of motion \cite {Riegert} (Efim Fradkin and Tseytlin \cite {FT},
built the same action in conformal gravity, practically at the same time \cite {Riegert}).
This construction was done for a particular class of constrained fields, and hence was regarded by Riegert 
as breaking diff-invariance. 

Komargodski and Schwimmer \cite {KS} showed that 
the same functional, $S_A$, but written in terms of unconstrained fields, gives a diff-invariant 
trace anomaly action.  This finding  placed the Riegert action on a solid
footing, which it lacked for many years. Moreover, 
ref. \cite  {GGTukhashvili}, showed that the Riegert action emerges as a local diff-invariant 
Wess-Zumino term in a coset for a non-linearly realized conformal symmetry, 
broken by the scale anomaly. 

These  findings offer an important perspective:  the GR action depends on the  metric $g$ and 
its derivatives.  The metric $g$ can formally 
be decomposed as $g=e^{2\s} {\bar g}$, and GR can be viewed as a theory 
non-linearly  realizing a Weyl symmetry that shifts $\s$ and conformally transforms 
$\g$, keeping $g$ invariant. In GR this is a "spurious" symmetry 
since the above split of $g$ is arbitrary, and the $\s$ field can be gauged away by 
the very same Weyl transformations. However, the Riegert action is a local diff-invariant functional  
of $\s$ and $\g$ which is not invariant under the "spurious" Weyl transformations \cite {KS}. 
Thus, the local diff-invariant action  containing the Einstein-Hilbert and Riegert terms, 
both written in terms of $\s$ and $\g$, can be viewed as an action non-linearly realizing 
the "spurious" Weyl symmetry  \cite {GGTukhashvili}.

The above arguments, however, suggests that something must be wrong: the field $\s$  
was "spurious" in GR, but becomes unremovable once GR is supplemented by the Riegert 
action. Indeed, in the GR action the metric field can absorb  the kinetic term of $\s$, 
rendering it in the Riegert action infinitely strongly coupled at arbitrarily low energies  \cite {Kurt} \footnote{Komargodski and 
Schwimmer constructed the local diff-invariant Riegert  action, and combined it with GR  
to prove the {\it a-theorem}  (for an earlier work using a 
non-local Riegert action for the a-theorem, see \cite {Anselmi}).  
They used the metric field for symmetry bookkeeping, but its dynamics 
was unimportant for the proof itself; hence the metric was frozen, 
and only the conformal mode (a dilaton) was utilized \cite {KS}. The issue discussed here  
does not affect the Komargodski-Schwimmer proof of the a-theorem, since their construction 
does not require a dynamical tensor field \cite {KS}. See more  comments in Section 4.}.  

One way to resolve this problem is to augment the classical GR action and only then couple it to a
quantum field theory. I will show that the following  action 
\beq
S_{R-\R}=M^2 \, \int d^4 x \sqrt{-g}\,R\,  -  {\bar M}^2 \, \int d^4 x \sqrt{-{\bar g}}\,{\bar R}\,,
\label{aGR}
\eeq
with ${\bar R}\equiv R(\g)$, $M=M_{\rm Pl}/\sqrt{2}>>{\bar M}$, and  the 
two metric tensors connected as
\beq
g_{\mu\nu} = e^{2\sigma} \, {\bar g}_{\mu\nu}\,,
\label{gsgbar}
\eeq
gives a viable low energy theory of gravity with a proper 
propagating conformal mode,
that can be consistently coupled to the Riegert action. 
The "spurious" Weyl transformation, $\s\to \s+\beta(x)$ , $\g \to e^{-2\beta(x)} \g$,  with 
an arbitrary $\beta(x)$, would have made $\s$ gauge-removable in GR, however, this is
 not a symmetry of the second term in the action (\ref {aGR}), and therefore $\s$ can't be gauged away. 
 Furthermore, the action (\ref {aGR}) possesses a global symmetry  that neither the first nor the 
second term on the r.h.s. separately has (Section 2). 

Combining (\ref {aGR}) with the results of \cite {Riegert,FT,KS}, the total  effective 
action  that captures the GR trace anomaly equation 
reads as follows:
\beq
S_{eff}=   S_{R-\R}\, + \, S_{A}(\sigma, {\bar g} )\,,
\label{S00}
\eeq
where $S_A$ has the form \cite {Riegert,FT,KS} 
\beq
S_{A} = a\int d^4x \sqrt {- {\bar g} } \left ( \sigma {\bar E} - 4 {\bar G}^{\mu\nu} {\bar \nabla}_\mu \sigma {\bar \nabla}_\nu \sigma  - 4 ({\bar \nabla}^2 \sigma) ({\bar \nabla} \sigma)^2 - 
2 ({\bar \nabla} \sigma )^4\right) + c\int d^4x \sqrt {- {\bar g}} \sigma {\bar W}^2,
\label{SA0}
\eeq
with the Euler (Gauss-Bonnet) invariant, ${\bar E}={\bar R}_{\mu\nu\alpha\beta}^2 - 4 {\bar R}_{\mu\nu}^2+{\bar R}^2$, and the Weyl tensor squared, ${\bar W}^2={\bar R}_{\mu\nu\alpha\beta}^2 - 2 {\bar R}_{\mu\nu}^2+{\bar R}^2/3$. 
The action (\ref {SA0})
emerges as a Wess-Zumino term in a $SO(2,4)/ISO(1,3)$ coset, 
which can be recast as a boundary term in 5D \cite {GGTukhashvili}; 
this distinguishes (\ref {SA0})  from other  terms 
in the effective field theory.  Note that the terms in $S_A$ belong to the general class of the Horndeski  theories 
giving rise to second order equations of motion \cite {Horndeski}.

There are an infinite number of additional higher dimensional counter-terms
supplementing  (\ref {S00}). All these terms will be suppressed by respective powers of 
the scale $M=\mpl/\sqrt{2}$, or of higher scales, such as $M(M/\M)^2$; they will be neglected.  
At the same time,  $S_A$ retains a finite number of terms which are 
suppressed by $\M<<M$, or by certain geometric mean scales  
such as $(\M M^2)^{1/3}$, all of them lower than $M$.

It is due to the separation of scales between $\M$ and $M$ that  one can regard  (\ref {S00}) as a  meaningful 
low energy action  obtained by  "integrating out"  quantum loops of a QFT. 
The coefficients $a$ and $c$ depend on numbers and representations of 
the low energy physical degrees of freedom \cite {Duff}. The  very same  
degrees of freedom, could also give rise to classical  sources for the gravitational field. Hence the action (\ref {S00}) should  
be supplemented by the classical action for the fields representing those low energy degrees of freedom, but 
without quantizing those fields  further\footnote{Not all quantum loops 
are proportional to the positive powers of $\hbar$ when massive fields are involved, 
some classical effects emerge from such loops 
\cite {Donoghue_hbar}. Hence, one might worry that without considering the quantum loops some classical 
effects would be lost.  However, keeping the respective classical fields in the effective 
action would enable one to retain those classical effects via nonlinear classical perturbation theory 
(or via exact classical or numerical solutions).}. 

The signature used in this work is "mostly plus", $(-,+++)$. The Ricci tensor 
convention is as follows, $R_{\mu\nu} = \partial_\alpha \Gamma^{\alpha}_{\mu\nu}-...$;
Riegert, following Ref. \cite {BirDav}, uses the "mostly minus" signature, 
and the opposite sign for the curvature tensor. Ref. \cite {KS} uses the 
"mostly minus" signature, the curvature convention opposite to Riegert's, and 
their $\tau$ and $g$ are, respectively, $-\s$ and $\g$ here.  The actions in refs. \cite {Riegert,FT}, \cite {KS}, and 
in eq. (\ref {SA0}) agree with each other after these different conventions 
are taken into consideration.

\section{The $R -\R$ theory}
\label{RminusRbar}

Consider the  action already quoted in the previous section
\beq
S_{R-\R}=M^2 \, \int d^4 x \sqrt{-g}\,R\,  -  {\bar M}^2 \, \int d^4 x \sqrt{-{\bar g}}\,{\bar R}\,,
\label{SRRbar}
\eeq
with the two metric tensors, $g$ and $\g$, related to one another by (\ref {gsgbar}). 
At the classical level, the above can be viewed as an action for 
one metric field -- say, the metric $g$ -- and for the scalar $\s$, which gets its 
proper-sign kinetic term from the $\R$ term\footnote{This would correspond to the Einstein frame. 
Section 4 will instead regard this action as a functional of $\g$ and $\s$, corresponding 
to the Jordan frame.}.

Let us consider  metric fluctuations above a flat background, $h_{\mu\nu} = g_{\mu\nu} -\eta_{\mu\nu}$, and 
decompose them in a  standard fashion according to the representations 
of the 3D rotation group (these  approximate well  fluctuations about an arbitrary nonsingular classical 
background at length scales much shorter than the characteristic curvature radius of the 
background):
\beq
h_{00} = 2n\,,~~ h_{0j} = v_j + \partial_j u\,,~~
h_{ij}= t_{ij} + \partial_i w_j + \partial_j w_i + 2\partial _i \partial_j \rho - 2 \delta_{ij} \tau\,,
\label{pert}
\eeq 
where $i,j=1,2,3$, $v_j$ and $w_j$ are transverse three-vectors, $t_{ij}$ is a transverse--traceless tensor, 
and $\tau$ is the conformal mode.  
Furthermore, one can choose the gauge, $u=0, \rho=0$. Then, the scalar part of the action 
(\ref {SRRbar}) -- which decouples from the tensor and vector parts -- equals to  the 
space-time integral of the following expression:
\beq
2M^2 \( -3 {\dot \tau}^2 + (\partial_j\tau)^2  -2 n \partial_j^2 \tau \) - 2{\M}^2 
\( -3( {\dot \tau} +\dot \sigma)^2 + (\partial_j(\tau+\sigma))^2  -2 (n+\sigma)  \partial_j^2 (\tau+\sigma) \)\,. 
\label{scalarL}
\eeq
If the  terms proportional to $\M^2$ were absent, the variation of  (\ref {scalarL}) w.r.t. $n$ would have given 
a constraint, $\partial_j^2\tau =0$,  rendering the conformal mode, $\tau$,  non-propagating. This 
however is no longer the case in (\ref {scalarL}): its variation w.r.t. $n$ gives another constraint,
 $\partial_j^2 \tau = \e^2 {\partial_j^2 \sigma}/(1-\e^2)$, that  relates 
the conformal mode $\tau$ to the scalar  $\sigma$, $\tau = \e^2 \sigma/(1-\e^2)$ (here, 
$\e=\M/M<<1$, and a zero mode of the Laplacian has been removed by choosing the appropriate  
spatial boundary conditions.) Substituting the latter into  (\ref {scalarL}), one gets 
\beq
- \,6 M^2 {(1-\e^2 )\over \e^2} \( \partial_\mu \tau \) \( \partial^\mu \tau \) = - \,6 {\M}^2 {1\over (1-\e^2)} \( \partial_\mu \sigma \)\( \partial^\mu \sigma \)\,,
\label{conf_mode}
\eeq
which is the kinetic term for the conformal mode with the proper sign. This is the key feature of the $R-\R$ theory.

It is convenient to  rewrite (\ref {SRRbar}) as follows:
\beq
S_{R-\R}= \int d^4 x \sqrt{-g}\,\(M^2\, R(g) - \Phi^2 \,R(g)\,  -  6\,  (\n \Phi)^2\)   \,,
\label{SRRbarPhi1}
\eeq
where $\Phi \equiv {\M} e^{-\s}$,  and the covariant derivative, $\n$, is that of the metric $g$. 
Owing to the choice of the sign of the second term on the right hand side of (\ref {SRRbar}),  
the kinetic term for the scalar $\Phi$ has the proper sign in (\ref {SRRbarPhi1}); this sign will remain the same  
after complete diagonalization  of the action (\ref {SRRbarPhi1}), as shown in Appendix.

The scale $\M$ enters the action (\ref {SRRbarPhi1}) only through the 
vacuum expectation value (VEV), $\langle  \Phi \rangle =\M$. Since $\M<< M$, 
this VEV is within the realm of the effective field theory.  The last two terms in 
 (\ref {SRRbarPhi1}) non-linearly realize a Weyl symmetry which transforms $g$, and is  
 explicitly broken by the  Einstein-Hilbert term.  
 
It is straightforward to check that the action (\ref {SRRbarPhi1}) is 
invariant w.r.t. the following transformations of the fields
\beq
{\Phi} \to {\Phi}\,+ {\omega \,M\,(1 -(\Phi/M)^2) \over 1+\omega\,\Phi/M},~~~
g_{\mu\nu} \to g_{\mu\nu}\,{(1+\omega \,\Phi/M)^2 \over 1-\omega^2}\,,
\label{symmetry0}
\eeq
where $\omega \equiv {\rm tanh} ( \lambda )$, with $\lambda$ 
being an arbitrary constant\footnote{The symmetry transformations (\ref {symmetry0}) may look mysterious, 
but their essence is simple: in  Appendix it is shown that certain non-linear conformal transformations of $g$ and $\Phi$, 
bring the action (\ref {SRRbarPhi1}) to that of GR coupled to a  massless scalar field kinetic 
term, which is invariant w.r.t. the shift symmetry.  
The shift transformation,  once rewritten in terms of the original variables,  gives 
(\ref {symmetry0}).}. 

Furthermore, there exists an invariant combination of the  metrics $g$ and $\g$  
\beq
{\hat g}_{\mu\nu} = g_{\mu\nu}  -  \e^2\,\g_{\mu\nu} = g_{\mu\nu}\,\(  1- {\Phi^2 \over M^2} \)\,,
\label{ghat}
\eeq
and if a QFT  coupled to $\hat g$ in a conventional manner, ${\cal L}_{\rm SM} = -{1\over 4} \sqrt {-\hg}\,\hg^{\mu\nu} 
\hg^{\alpha \beta} F_{\mu \alpha} F_{\nu \beta} + \cdots$, then it will also be invariant.
Since one should require, $\e\equiv (\M/M)<<1$, the coupling to the QFT is  
approximated by the coupling to the metric $g_{\mu\nu}$. The resulting classical equations 
approximate GR, as long as $\M<<M$ \footnote{
One could consider a different approach in which 
QFT fields would couple to $g$, instead of $\hat g$, thus breaking the symmetry (\ref {symmetry0}).
This would result in new terms generated in the effective action due to the QFT loops,
notably  the mass term for the $\sigma$ field would be induced. The latter would be
proportional to some positive power of the UV  scale of the QFT, suppressed by powers of $M$. 
Furthermore, $\sigma$ would couple to the stress-tensor in the linearized approximation,
providing a gravity-competing force at distances smaller than inverse mass of $\sigma$. 
One would then need to impose a constraint on $\M$ to suppress the coupling of $\sigma$ to 
the stress-tensor. While this is a logical possibility, it would also lead to the trace 
anomaly equation being modified as compared with that of GR (such modifications can be 
kept small by imposing constraints on  $\M$). 
The present work focuses  on a scenario where the symmetry (\ref {symmetry0}) 
is preserved by the QFT coupling, but can be adopted
to the case when the matter couples to $g$.}.

The requirement of the invariance under (\ref {symmetry0}) 
prohibits certain  terms to be added to the action  (\ref {SRRbarPhi1}).
For instance, an additional kinetic term for $\Phi$, the cosmological constant, $\sqrt{-g}$, 
the quadratic mass term, $\sqrt{-g}\Phi^2$, and the 
quartic term, $\sqrt{-g} \Phi^4$, don't respect the symmetry (\ref {symmetry0}), and are prohibited 
to enter the action with arbitrary coefficients. That said, there is a particular combination of the latter three 
terms which is invariant under (\ref {symmetry0})
\beq 
\pm\,\Lambda^4 \sqrt{-g}\, \(1- {\Phi^2 \over M^2} \)^2\,,
\label{addCC}
\eeq
with $\Lambda$ being some arbitrary dimensionful constant. 
If the above term is included in the action then both the quadratic and 
quartic terms for $\Phi$ will  be connected to the cosmological constant. 
The cosmological constant $\Lambda$ will be tuned to zero (or be vanishingly  
small as compared to $\M$) and this will also nullify the quadratic and quartic terms in 
(\ref {addCC}).

The term in (\ref {addCC}) is nothing other than $\pm \Lambda^4 \sqrt{-\hat g}$. 
One could also consider the terms, $ \sqrt{-\hat g}\,{\hat R}$, and 
$  \sqrt{-\hg} {(\hn \Phi )^2} /(1-\Phi^2/M^2)^2 $,  to be added to the action  (\ref {SRRbarPhi1}), 
with arbitrary coefficients; these terms would individually respect the symmetry (\ref {symmetry0}). 
However, only one linear combination of the above two terms retains the structure of the action 
(\ref {aGR}) intact, which is necessary to preserve the same 
trace anomaly equation as one gets in GR (see Section 4). Therefore, in addition to 
the fine tuning of the cosmological constant, one needs to adopt another fine tuning of 
the relative coefficients between the above two terms\footnote{One could relax this tuning somewhat 
to obtain the trace anomaly equation that would slightly differ from that of GR; this is easy to do
but there is no urgency to pursue such extensions.}. 

The action (\ref {aGR}), and the relation (\ref {gsgbar}),  are invariant under the following 
"duality" transformations: $M \leftrightarrow \M$, $g\leftrightarrow -\g$, $\s\rightarrow -\s$.
The latter leads to $\hg \rightarrow \e^{-2} \hg$,   rendering the following three terms 
invariant, $M^4 \sqrt{-\hat g}$, $M^2\sqrt{-\hat g}\,{\hat R} $ and 
$\sqrt{-\hg} {(\hn \Phi )^2}/ (1-\Phi^2/M^2)^2 $; furthermore, coupling of $\hg$ to 
massless scalars, spinors, and vector fields can be made invariant by inserting appropriate 
powers of $M$ in front of their kinetic terms. Therefore, the above "duality symmetry" 
does not help to avoid the need for the fine tuning of the two parameters discussed above.

The equations of motion that follow from (\ref {SRRbarPhi1}) read:
\beq
(M^2 &-&\Phi^2) \( R_{\mu\nu} -{1\over 2} g_{\mu\nu} R  \) + \(\nabla_\mu \nabla_\nu - g_{\mu\nu} \nabla^2 \) \Phi^2 =
6 \(\nabla_\mu \Phi \nabla_\nu\Phi - {1\over 2} g_{\mu\nu} (\nabla \Phi)^2 \),\nonumber \\
&-&\Phi R + 6 \nabla^2 \Phi=0\,,
\label{EOM0}
\eeq
where one can easily include the matter stress-tensor on the r.h.s. of  the first equation.
It is straightforward to show that a flat FLRW solution of  GR is also a solution of (\ref {EOM0}), 
amended by the matter stress-tensor.

\vspace{0.1in}

\section{Toward the quantum effective theory}
\label{QEFT}

In general, one could quantize (\ref {SRRbarPhi1}) as an effective low energy 
action (see \cite {Donoghue}, and references to earlier literature). 
However, there is no need for a general approach here. 
Instead,  one could regard gravity as a dynamical classical field 
coupled to a QFT \cite {BirDav}.  This is justified at  energy 
scales much below the Planck scale.

Quantum loops of the QFT, including the loops by which one would integrate out possible 
Planck scale physics, will generate an infinite number of higher-dimensional terms that one 
needs to include in the effective action. To deal with the loop divergencies one could use, e.g.,  
dimensional regularization. 

Due to the symmetry (\ref {symmetry0}) of the action (\ref {SRRbarPhi1}) and of its coupling to a QFT, 
all the higher dimensional terms should also be invariant under (\ref {symmetry0})\footnote{I will neglect 
non-perturbative quantum gravity effects due to,  e.g.,  black holes or worm holes,  which 
are expected to violate global symmetries, and in particular to violate  (\ref {symmetry0}). Such violations are 
exponentially suppressed  at low energies in the quasi-classical approximation \cite {SusskindLinde}; 
they should be contrasted to the violation due to the trace anomaly, which is 
suppressed only by the powers of the scale $\M$, and 
will be included in the effective action.}.  
Hence, one could identify all the symmetry preserving terms by writing all possible invariants 
in terms of the metric $\hat g$, expressed in terms of $g$ and $\Phi$. 
Furthermore, the metric $\hat g$ approximately equals to $g$, up to the corrections of the order 
of $1/M^2$, hence,   one could just use $g$  as an approximation. The goal of this section is to 
understand at what energy scales are those counter-terms significant. 

To achieve this goal, let us  introduce dimensionful fields 
\beq
{H}_{\mu\nu} = {M\( g_{\mu\nu} -\eta_{\mu\nu} \)}, ~~~~ {\Sigma} = {\M} -\Phi\,,
\label{firlds}
\eeq
and consider two different limits of the theory (\ref {SRRbarPhi1}).

First, consider the limit $M\to \infty$, while $\M$ is fixed, 
and ${H}$ and $\Sigma$ are finite. It is then straightforward to see that the Lagrangian 
in (\ref {SRRbarPhi1}) reduces to a free field theory of ${H}$ and $\Sigma$
\beq
-{1\over 4}\, {H}^{\mu\nu} G_{\mu\nu} ({H} ) - 6 (\partial {\Sigma})^2\,,
\label{free1}
\eeq
where $G_{\mu\nu} ({H} )$ denotes the linearized Einstein tensor. The coupling to the stress-tensor, 
$T_{\mu\nu}$, is proportional to $ {H}_{\mu\nu}T^{\mu\nu}/M$, up to additional corrections  of the order 
of ${\cal O} \(\M^2/M^3 \)$, and also vanishes in the limit as long as 
the stress-tensor is held finite. These considerations show that the loop-generated 
counter-terms in the full nonlinear theory should vanish in the  $M\to \infty$ limit.

Second, consider the limit $M\to \infty$, $\M\to \infty $, with $\e =(\M/M)<<1$ being fixed;
the fields ${H}$ and $\Sigma$ are held finite in this limit, too. The resulting Lagrangian 
reads
\beq
-{1-\e^2\over 4} \,{H}^{\mu\nu} G_{\mu\nu} ({H} ) + 2 \e \,{ \Sigma} \,R_L({H})- 6 (\partial { \Sigma})^2\,,
\label{free2}
\eeq
where $R_L ({H} )$ denotes the linearized Ricci scalar, and  the coupling to the matter stress-tensor
is proportional to $ {H}_{\mu\nu}T^{\mu\nu}/M$,  up to the additional corrections  of the order 
of ${\cal O} \(\M^2/M^3 \)$. The expression (\ref {free2})  represents a 
free field theory of a kinetically mixed tensor and scalar fields, and can easily be diagonalized by a 
linear conformal transformation of $ {H}_{\mu\nu}$.  Therefore,  the counter terms of the 
full nonlinear theory ought to vanish in the second limit, too. The latter 
condition is  more restrictive than the one obtained from the first limit above. 
It implies that  there will not exist counter-terms proportional to $\M^{p}/M^q$, with $p\ge q$. In particular, 
the following counter terms  
\beq
\M^2 \({\M\over M}\)^n\sqrt{-g}\,{ \Sigma}^2, ~~~ \({\M\over M}\)^m\sqrt {-g}\,{\Sigma}^4,
\label{relevant}
\eeq
will be absent for arbitrary integers $n\ge 0$ and $m\ge 0$  (i.e., 
such terms will not be generated with $n\ge 0$ and $m\ge 0$ even if they form 
combinations that are invariant under (\ref {symmetry0})).

Let us now summarize the results of the above discussions in terms of
the fields of  the action (\ref {SRRbarPhi1}).  One concludes that the counter-terms 
in  (\ref {SRRbarPhi1})  will be  proportional to
\beq
{R^3\over M^2}, ~~{R\square R \over M^2}, ~~{\Sigma^2 R^2 \over M^2}, ~~{\Sigma^2 R\square R \over M^4}\,,...
\label{counter}
\eeq
Each of these counter-terms will have corrections that make them invariant under (\ref {symmetry0}),
however,  each of these corrections are higher order in $1/M^2$.

All of these counter terms are  suppressed by  $M$. Furthermore, due to the VEV, $\langle \Phi \rangle =\M$,
some of the higher dimensional operators will end up depending on $\M$, too:
\beq
{\M^2 R^2 \over M^2}, ~~{\M^2 R\square R \over M^4}\,...
\label{counter2}
\eeq
The lowest  scale that suppresses the latter operators is $M (M/\M)$, which is significantly higher than
$M$, thanks to  $\M<<M$. 

All the arguments above apply to conventional local
counter-terms encountered in  the conventional perturbative series expansion
in powers of $\n^2/M^2$ \cite {Donoghue}.  
This however says nothing about possible non-local terms that may arise from the loops. Due to their 
nonlocal nature such terms could remain significant even when $\n^2<<M^2$. 
The trace anomaly 
introduces precisely such terms, which can then be rewritten as local term but at the expense 
of introducing a new field, $\s$.  The new terms in the effective action are suppressed 
by the scale much smaller than $M$. Hence, it  is meaningful to keep the trace-anomaly 
induced terms, but ignore all the other conventional higher dimensional counter terms suppressed by 
$M$ and higher scales, as done below.

\section{The effective action}

The scale anomaly in the trace of a massless QFT  stress-tensor reads \cite {Duff}:
\beq
 T^\mu_\mu =  a E(g) + c W^2(g) + b \square R(g)\,,
\label{Trace0}
\eeq
where the coefficients $a$ and $c$  depend on the  field content of the theory, 
while $b$ is arbitrary, and for that reason will not be included\footnote{Both, 
$b$, and the gauge field trace anomaly \cite {Crewther,ChanowitzEllis}, 
can easily be included in the 
effective action \cite{Riegert}.}. The goal is to find an effective action, ${\tilde S}_A(g)$, 
which incorporates the loop effects of the QFT so that 
the trace of its metric variation gives (\ref {Trace0})
\beq
 \,T^\mu_\mu =  \,g^{\alpha \beta}{2 \over \sqrt {-g}} \,{\delta {\tilde S}_A(g) \over \delta g^{\alpha\beta} }\,=
 a E(g) + c W^2(g)\,.
 \label{Var_Trace}
\eeq
Riegert \cite {Riegert}, argued that  ${\tilde S}_A(g)$ 
cannot be written as a local diff-invariant functional if one uses only one field, $g$.
Yet, the variation of the total action, $S_{GR}(g)+{\tilde S}_A(g)$, should give rise 
to a local equation for the trace anomaly, with the r.h.s. defined by (\ref {Var_Trace}). 
 
To find the action, Riegert introduced a decomposition,  $g_{\mu\nu} = e^{2\s}\,{\g}_{\mu\nu}$, 
with the metric $\g$ restricted to have a fixed determinant \cite {Riegert}. Using this decomposition, 
Riegert constructed a local action, $S_A(\g, \s)$, such that the variation of $S_{GR}(e^{2\s}\g)+{S}_A(\g,\s)$ w.r.t. $\s$ 
gives the trace anomaly equation. The functional $S_A(\g, \s) $ was regarded in \cite {Riegert}
as breaking diff-invariance since the determinant of $\g$ had been fixed  \cite {Riegert}. 

It is however straightforward to restore diff-invariance in  $S_A(\g, \s)$: one could use 
exactly the same action, $S_A(\g, \s)$, but declare that the restriction on the determinant of $\g$ has been
lifted. Conversely,  such a diff-invariant 
$S_A(\g, \s)$ would yield back Riegert's original local but non-invariant
$S_A(\g, \s)$, after gauge-fixing the determinant of $\g$. 
 
Logically, Komargodski and Schwimmer \cite {KS} reconstructed 
Riegert's action $S_A(\g, \s)$, as a 
functional of  two fields, $\s$ and $\g$, without assuming any constraint on $\g$, 
only requiring that the action reproduce the trace anomaly upon simultaneous Weyl  
transformations of $\s$ and $\g$. 

Furthermore, the same $S_A(\g, \s)$ was obtained 
as a Wess-Zumino term in a coset construction for non-linearly realized 
conformal symmetry \cite {GGTukhashvili}.

The above  facts suggest that the local and invariant Riegert action, $S_A(\g, \s)$,
given in  (\ref {SA0}), should perhaps play a more fundamental role than it did in 
Riegert's construction, where it was merely  used as an intermediate step toward motivating 
a non-local action\footnote{Riegert's subsequent 
procedure of constructing  a non-local but 
diff-invariant functional introduces a four derivative term in the action \cite {Riegert}.
This rout unavoidably leads to a negative energy state, a ghost (or to violation to unitarity). 
It is also an unnecessary rout --  as argued,  a local 
and diff-invariant effective action for the trace anomaly does exist. More general 
actions with four-derivative terms were explored in \cite {Motolla}, 
with some potentially interesting applications. Regretfully, 
these actions also have  ghosts.}.

To this end, one could add the action (\ref {SA0}) to the GR
action expressed in terms of $\s$ and $\g$, $S_{GR}(e^{2\s}\g)$, and treat both  
$\s$ and $\g$ as dynamical fields without assuming any constraint on $\g$.  
This theory, however, describes an infinitely strongly 
coupled system when restricted to its Minkowski background \cite {Kurt}.  
It is a strongly coupled theory with an unacceptably  low strong coupling scale on 
any small curvature background.  This is so because the theory,  $S_{GR}(e^{2\s} \g)+{S}_A(\g,\s)$, 
does not contain a kinetic term for $\s$.  An apparent kinetic term for $\s$ in  $S_{GR}(e^{2\s} \g)$ 
can be removed by a field redefinition of the metric, $\g = e^{-2\s} g$, rendering the $\s$ field endowed 
with nonlinear interactions in ${S}_A(g,\s)$, but without  a quadratic kinetic term. 
Such a theory is intractable as an effective field theory\footnote{This is not an issue 
for the work \cite {KS} insofar as $\g$  is not a dynamical field and is frozen 
to equal to the flat space metric; in the latter case the GR action gives a kinetic term for $\s$, albeit 
with a wrong overall sign. This sign can be flipped to the correct one by choosing 
a "wrong sign" GR term to start with; this  does not cause a problem in \cite {KS} since the 
theory  does not use the dynamics of tensor perturbations.}.

To bring some normalcy, one needs to introduce 
a kinetic term for $\s$. However, adding in any  
new term explicitly depending on  $\sigma$  -- besides the ones 
already present in (\ref {SA0}) -- would spoil the recovery of the correct trace anomaly. 
One way out is to use the $R-\bar R$ theory 
as a functional of $\g$ and $\s$, where  the new term, $\bar R$, does not depend on $\s$.

This leads one to explore a new action consisting of the $R-\bar R$ theory plus (\ref {SA0}):
\beq
S_{ eff}(\g, \s) =M^2 \, \int d^4 x \sqrt{-{\g}}\,  e^{2\s}   \(    {R}({\g}) +6\, ({\bar \n} \s)^2  \)  -  {\bar M}^2 \, \int d^4 x \sqrt{-{\bar g}}\,R({\bar g}) +\nonumber \\ 
a\int d^4x \sqrt {- {\bar g} } \left (\sigma  {\bar E} - 4 {\bar G}^{\mu\nu} {\bar \nabla}_\mu \sigma {\bar \nabla}_\nu \sigma  - 4 ({\bar \nabla}^2 \sigma) ({\bar \nabla} \sigma)^2 - 
2 ({\bar \nabla} \sigma )^4\right) + c\int d^4x \sqrt {- {\bar g}} \,\sigma\, {\bar W}^2\,.
\label{SRRbarSA}
\eeq
This is a local diff-invariant functional 
of $\s$ and $\g$. It can be viewed as an effective action for 
quantized  $\s$ and  QFT \footnote{Note that the Riegert action in the second line in 
(\ref {SRRbarSA})  contains the Galileon terms that are not 
suppressed by $M$. These can be understood as terms 
obtained by "integrating in" the $\sigma$ field to make 
the otherwise nonlocal anomaly action be local. In such a case,  not all the terms containing $\sigma$ 
should be suppressed by $M$. Conversely, integrating out $\sigma$ would lead to nonlocal  terms,  all of them 
suppressed by $M$.}.

The classical matter fields -- not shown in  (\ref {SRRbarSA})   -- 
couple to the metric ${\hat g}_{\mu\nu} = g_{\mu\nu} (1- \M^2 e^{-2\s}/M^2) \simeq g_{\mu\nu}= e^{2\s} \g_{\mu\nu}$. 
Variation of the action (\ref {SRRbarSA}) with respect to $\s$, with a subsequent substitution
of $\g = e^{-2\s} g$, gives an equation that does not depend on $\s$
\beq
{\delta S_{eff}(\sigma, {\bar g}) \over \delta \sigma}|_{ {\bar g}_{\mu\nu} 
= e^{-2\sigma} \, {g}_{\mu\nu}\ }=0\,, ~\Rightarrow~
2M^2 R = aE(g) + cW^2(g)\,,
\label{AnomEq}
\eeq
which is exactly the trace-anomaly equation.

Now that the tensor $\g$ is an unconstrained dynamical field, 
one should vary (\ref {SRRbarSA}) w.r.t. $\g$, too. 
This variation will give ten modified Einstein's equations.
One can show that the trace of these equations does not coincide 
with the equation (\ref {AnomEq}). Thus, there are eleven independent equations: 
one equation of motion of $\s$, and ten equations 
for ten components of the symmetric tensor $\g$. By determining $\s$ 
and $\g$ from these equations, one can determine ${\hat g}\simeq g$, which is the 
metric experienced by the classical matter fields. The modified Einstein equations 
approximate well the conventional equations as long as $\M<<M$, and
as long as the stress-tensor is much smaller than $\M^4$. 

Alternatively, one could rewrite the ten equations for $\g$ 
as ten equations for $g$, which would also depend on $\s$. 
The $\s$ equation, on the other hand, can entirely be 
kept in terms of $g$ (\ref {AnomEq}).
Thus, there would still be eleven equations for eleven 
unknowns, $g$ and $\sigma$.

The theory given by (\ref {SRRbarSA}) is strongly coupled at the scale $\M$.
This can be seen in the flat space expansion of  (\ref {SRRbarSA})
obtained either by taking the limit $M \to \infty$, or 
by  the substitution  
\beq
g_{\mu\nu} = \eta_{\mu\nu} \(  1- {\Phi^2/ M^2} \)^{-1},~~~
\g_{\mu\nu} = e^{-2\s}\,\(  1- {\Phi^2/M^2} \)^{-1} \eta_{\mu\nu}.
\label{sub}
\eeq
In either case, the theory 
reduces to a conformal Galileon of the canonically normalized field, 
${\pi} \equiv \sigma\M$\cite {Rattazzi}, with the nonlinear 
Galileon terms suppressed by the scale $\M$
\beq
S_{eff}|_{\g_{\mu\nu} \simeq e^{-2\pi/\M} \eta_{\mu\nu}} \simeq \int d^4 x \( 
-6 e^{-2\pi/\M} (\partial \pi )^2 + 4a \( -{ {\partial^2 \pi} (\partial \pi)^2 \over \M^3} +   
{(\partial \pi)^2(\partial \pi)^2 \over 2\M^4}\) +\dots \),
\label{confGal}
\eeq
where the dots stand for sub-leading terms suppressed by powers of $M$. 

Below the scale $\M$ the full 
theory (\ref {SRRbarSA}) is weakly coupled and describes three degrees of freedom --  
two helicity states of a massless graviton, and one massless scalar, ${\pi} \equiv \sigma\M$.
The later becomes strongly coupled at the scale of the order of 
$\M$, as seen from the Galileon terms in (\ref {confGal}). 
The Lagrangian (\ref {confGal}) makes it clear  that the theory without the $\M^2 \sqrt{-\g}{\bar R}$ term is 
untenable  -- taking $\M \to 0$ leaves the nonlinear terms in (\ref {confGal}) infinitely strongly coupled.

The additional scalar, $\s$, does not couple to the stress tensor  in the linearized 
approximation in  Minkowski space, therefore there is no fifth force constraint. 
It can couple to matter on curved backgrounds. Physical consequences of the 
action (\ref {SRRbarSA})  will be studies elsewhere.
\vspace {0.1in}

{\bf Acknowledgments:} I'd like to thank Massimo Porrati, David Spergel, and Giorgi Tukhashvili  
for useful discussions, and Pedro Fernandes and Arkady Tseytlin for helpful communications. 
The work was supported in part by NSF grant PHY-2210349.

\vspace {0.2in}

{\bf Comments added:}  After this article appeared in the arXiv, Pedro Fernandes 
communicated  that the action which he derived  in \cite{PF} using different principles, 
is equivalent to the action given above in  (\ref {SRRbarSA}) (up to some obvious terms).  
Indeed, if one transforms the action (\ref {SRRbarSA})
from the Jordan frame to the Einstein frame, and then changes the 
field $\sigma$ to $-\phi$, one obtains the respective terms in the action  
(14)  of \cite {PF}.  Moreover, the "duality" discussed in the present work 
in Section 2, transforms the action (\ref {SRRbarSA})  
into the action (14) of ref.  \cite {PF}. 

The action in ref. \cite {PF} was derived based on purely classical principles,
by answering the question: what is the most general scalar-tensor theory 
of gravity with the second order equations of motion that has the   
scalar equation invariant under the conformal transformations, 
even if the action is not conformally invariant? Furthermore, the  invariance 
of the scalar equation of motion enabled Fernandes to find exact 
analytic solutions of the theory \cite {PF}.

It is encouraging that such a different  principle gave rise to 
the action of \cite {PF}, which was derived in the present work based on  
the consistency of the effective field theory of gravity with the quantum trace anomaly.
The anomaly perspective of the present work is essential: its 
outcome is that GR needs to be amended by a  scalar degree of freedom --  as it is done in the trace 
anomaly action (\ref {SRRbarSA}) -- in order for GR to be consistently 
coupled  to a QFT with the trace anomaly. If so, then  
there has to exist a new physical scale, $\M << \mpl$, 
governing the interactions of the scalar with itself,  and  governing -- along with $\mpl$ -- 
its interactions with the SM fields.  An alternative to the above is to cancel 
the trace anomaly by introducing other light degrees of freedom.

\subsection*{Appendix}

A complete  diagonalization of  (\ref {SRRbarPhi1}) can be done by using the following conformal transformation  
\beq
g_{\alpha\beta} = {\hg}_{\alpha\beta}\,\(  1- {\Phi^2\over M^2} \)^{-1}\,.
\label{gghatPhi}
\eeq 
It brings the action for the metric $\hg$ to the canonical form
\beq
S_{R-\R}=  \int d^4 x \sqrt{-\hg}\, \( M^2 \, R(\hg)\,  -   \, 
 { 6\,{(\hn \Phi )^2} \over (1-\Phi^2/M^2)^2 } \) \,.
\label{SRRbar_sigma}
\eeq
The above conformal transformation is non-singular, and 
(\ref {SRRbar_sigma})  is sensible, as long as 
\beq
\Phi =\M e^{-\sigma} << M, ~~~ \Rightarrow~~~ \sigma >>  -\,ln (M/\M)\,.
\label{validity1}
\eeq  
In addition, the effective theory is valid if $|\s|<<1$, which is a stronger constraint.

The scalar field action in (\ref {SRRbar_sigma}) can be simplified further: it 
can be reduced to a free field interacting with gravity, thanks to the following non-linear field redefinition
\beq
U= \( {1+\Phi/M \over 1-\Phi/M} \)^{1/2}\,.
\label{V}
\eeq
The resulting action reads:
\beq
S_{R-\R}=M^2 \,   \int d^4 x \sqrt{-\hg}\,\( R(\hg)\,  -   \,  6 \, (U^{-1} \hn U )^2   \) \,.
\label{SRRbarV}
\eeq
The latter makes the symmetries of the sigma model more explicit: the Lagrangian is invariant 
w.r.t. the  rescaling 
\beq
U\,\to e^{\lambda }\, U\,,
\label{symmetry}
\eeq
where $\lambda$ is an arbitrary constant. This symmetry is the shift symmetry of  
the scalar field  ${\rm ln}(U)$ that has only a kinetic 
term  in (\ref {SRRbarV}).

Note that $\hat g$ and $U$ are related to $\g$ and $\s$ through nonlinear transformations.
However, the path integral  $Z(\hat g)$  does not equal to the path integral $Z(g)$. 
In other words, the nonlinear conformal transformation from $\g$ to $\hat g$, and the 
quantization procedure for the scalar and QFT, do not commute 
with one another. If one were to start with (\ref {SRRbarV}) and combine it with the Riegert  
action for $\hat g$, one would get an infinitely strongly coupled theory. Instead, 
the Riegert action that needs to be added to (\ref {SRRbarV}) can be obtained from the 
Riegert action for $\g$, by using  the respective nonlinear conformal transformation 
from $\g$ to $\hat g$. The obtained action will be strongly coupled at the scale $\M$.



\end{document}